\documentclass[aps,prl,preprint,tightenlines,superscriptaddress,showpacs,byrevtex,subfigure]{revtex4}

\usepackage{graphicx}
\usepackage{color}

\begin{document}
\begin{flushright}
KEK Preprint 2008-37\\
Belle Preprint 2008-27\\
NTLP  Preprint 2008-01
\end{flushright}
\title{ \quad\\[0.5cm] 
Search for Lepton-Flavor-Violating $\tau$ 
{Decays}\\
into a Lepton and an $f_0(980)$ Meson
}

\begin{abstract}
We search for lepton-flavor-violating $\tau$ decays 
into a lepton (electron or muon) and {an} $f_0(980)$ meson
using
671 fb$^{-1}$ of data collected 
with the Belle detector at the 
KEKB asymmetric-energy $e^+e^-$ collider. 
{No events are observed and we set the 
following 90\% C.L. upper limits on the branching fraction 
{products:}}
${\cal{B}}(\tau^-\to e^-f_0(980))
\times
{\cal{B}}(f_0(980)\to\pi^+\pi^-)<3.2\times 10^{-8}$ 
and
${\cal{B}}(\tau^-\to\mu^-f_0(980))
\times
{\cal{B}}(f_0(980)\to\pi^+\pi^-)<3.4\times 10^{-8}$.
This is the first search {performed} 
for these modes.
\end{abstract}
\affiliation{Budker Institute of Nuclear Physics, Novosibirsk, Russia}
\affiliation{Chiba University, Chiba, Japan}
\affiliation{University of Cincinnati, Cincinnati, OH, USA}
\affiliation{The Graduate University for Advanced Studies, Hayama, Japan}
\affiliation{Hanyang University, Seoul, South Korea}
\affiliation{University of Hawaii, Honolulu, HI, USA}
\affiliation{High Energy Accelerator Research Organization (KEK), Tsukuba, Japan}
\affiliation{Institute of High Energy Physics, Chinese Academy of Sciences, Beijing, PR China}
\affiliation{Institute for High Energy Physics, Protvino, Russia}
\affiliation{Institute of High Energy Physics, Vienna, Austria}
\affiliation{Institute for Theoretical and Experimental Physics, Moscow, Russia}
\affiliation{J. Stefan Institute, Ljubljana, Slovenia}
\affiliation{Kanagawa University, Yokohama, Japan}
\affiliation{Korea University, Seoul, South Korea}
\affiliation{Kyungpook National University, Taegu, South Korea}
\affiliation{\'Ecole Polytechnique F\'ed\'erale de Lausanne, EPFL, Lausanne, Switzerland}
\affiliation{Faculty of Mathematics and Physics, University of Ljubljana, Ljubljana, Slovenia}
\affiliation{University of Maribor, Maribor, Slovenia}
\affiliation{University of Melbourne, Victoria, Australia}
\affiliation{Nagoya University, Nagoya, Japan}
\affiliation{Nara Women's University, Nara, Japan}
\affiliation{National Central University, Chung-li, Taiwan}
\affiliation{National United University, Miao Li, Taiwan}
\affiliation{Department of Physics, National Taiwan University, Taipei, Taiwan}
\affiliation{H. Niewodniczanski Institute of Nuclear Physics, Krakow, Poland}
\affiliation{Nippon Dental University, Niigata, Japan}
\affiliation{Niigata University, Niigata, Japan}
\affiliation{University of Nova Gorica, Nova Gorica, Slovenia}
\affiliation{Osaka City University, Osaka, Japan}
\affiliation{Osaka University, Osaka, Japan}
\affiliation{Panjab University, Chandigarh, India}
\affiliation{Saga University, Saga, Japan}
\affiliation{University of Science and Technology of China, Hefei, PR China}
\affiliation{Sungkyunkwan University, Suwon, South Korea}
\affiliation{University of Sydney, Sydney, NSW, Australia}
\affiliation{Toho University, Funabashi, Japan}
\affiliation{Tohoku Gakuin University, Tagajo, Japan}
\affiliation{Tohoku University, Sendai, Japan}
\affiliation{Department of Physics, University of Tokyo, Tokyo, Japan}
\affiliation{Tokyo Metropolitan University, Tokyo, Japan}
\affiliation{Tokyo University of Agriculture and Technology, Tokyo, Japan}
\affiliation{IPNAS, Virginia Polytechnic Institute and State University, Blacksburg, VA, USA}
\affiliation{Yonsei University, Seoul, South Korea}
\author{Y.~Miyazaki} 
\affiliation{Nagoya University, Nagoya, Japan}
\author{I.~Adachi} 
\affiliation{High Energy Accelerator Research Organization (KEK), Tsukuba, Japan}
\author{H.~Aihara} 
\affiliation{Department of Physics, University of Tokyo, Tokyo, Japan}
\author{K.~Arinstein} 
\affiliation{Budker Institute of Nuclear Physics, Novosibirsk, Russia}
\author{T.~Aushev} 
\affiliation{\'Ecole Polytechnique F\'ed\'erale de Lausanne, EPFL, Lausanne, Switzerland}
\affiliation{Institute for Theoretical and Experimental Physics, Moscow, Russia}
\author{A.~M.~Bakich} 
\affiliation{University of Sydney, Sydney, NSW, Australia}
\author{A.~Bay} 
\affiliation{\'Ecole Polytechnique F\'ed\'erale de Lausanne, EPFL, Lausanne, Switzerland}
\author{I.~Bedny} 
\affiliation{Budker Institute of Nuclear Physics, Novosibirsk, Russia}
\author{U.~Bitenc} 
\affiliation{J. Stefan Institute, Ljubljana, Slovenia}
\author{A.~Bondar} 
\affiliation{Budker Institute of Nuclear Physics, Novosibirsk, Russia}
% ...
\collaboration{The Belle Collaboration}
\noaffiliation

\maketitle
 \section{Introduction}

{Lepton flavor violation (LFV)
in charged lepton decays is forbidden 
in  the Standard Model  
{and remains} 
highly suppressed 
even if neutrino mixing is included.}
However, LFV 
{is expected} in various extensions of the 
Standard Model,
such as supersymmetry, leptoquark and
{many other models, e.g., 
various MSSM scenarios consider $\tau^-$
lepton decays into semileptonic final states~\cite{brig,arg}.}
If LFV occurs via {a} Higgs mediated LFV mechanism, 
{$\tau^-$ leptons can decay
into $\ell^- f_0(980)$, 
where $\ell^-$ is either an electron or a muon,
through a scalar Higgs 
boson~\cite{lf0}\footnotemark[2].}
{The decays} $\ell^- \pi^0$, 
$\ell^- \eta$ and $\ell^-\eta'$ 
{are mediated by} a pseudoscalar Higgs boson
{while} $\ell^-\mu^+\mu^-$ 
{can be mediated through both} scalar and pseudoscalar Higgs bosons~\cite{higgsLFV}.
Searches for $\tau^-\to\ell^-f_0(980)$ have not yet been 
{reported}
{while 
{there are already}
upper limits on
the branching fractions for $\tau^-\to\ell^-\pi^0~(\eta,~\eta')$ and 
$\tau^-\to\ell^-\mu^+\mu^-$ 
at the ${\cal{O}}(10^{-8})$ level~\cite{leta,lll}.}
With certain combinations of 
new {physics parameters} 
the branching fractions for 
$\tau^-\to\ell^-f_0(980)$ decays can be as  high as $10^{-7}$,
which is already accessible {in} high-statistics  
{$B$-factory} experiments.

Here we report our {search} for $\tau^-\rightarrow\ell^{-}f_0(980)$ 
with a data sample of 671 fb$^{-1}$ collected at the $\Upsilon(4S)$ resonance
and 60 MeV below it with the Belle detector at the KEKB 
asymmetric-energy   $e^+e^-$ collider~\cite{kekb}.

\footnotetext[2]{{Throughout this paper,
charge-conjugate modes are {implied} unless stated  otherwise.}}

The Belle detector is a large-solid-angle magnetic spectrometer that
consists of a silicon vertex detector (SVD),
a 50-layer central drift chamber (CDC),
an array of aerogel threshold Cherenkov counters (ACC), 
a barrel-like arrangement of
time-of-flight scintillation counters (TOF), 
and an electromagnetic calorimeter {(ECL)}
comprised of CsI(Tl) 
{crystals,} all located inside
a superconducting solenoid coil
that provides a 1.5~T magnetic field.
An iron flux-return located {outside} the coil is instrumented to 
detect $K_{\it{L}}^0$ mesons
and to identify muons (KLM).
The detector is described in detail elsewhere~\cite{Belle}.

We use particle identification
likelihood variables based on
the ratio of the energy
deposited in the 
ECL to the momentum measured in the SVD and CDC,
the shower shape in the ECL,
the particle range in the KLM,
the hit information from the ACC,
{the $dE/dx$ information in the CDC,}
and {the {particle} time-of-flight} from the TOF.
To distinguish hadron species,
we use likelihood ratios,
for instance, 
${\cal{P}}(i/j) = {\cal{L}}_i/({\cal{L}}_i + {\cal{L}}_{j})$,
where ${\cal{L}}_{i}$ (${\cal{L}}_{j}$)
is the likelihood for the detector response 
to the track with flavor hypothesis $i$ ($j$). 
{{For lepton identification,
we {form} likelihood ratios ${\cal P}(e)$~\cite{EID} 
and ${\cal P}({\mu})$~\cite{MUID}
based on the}
electron and  muon probabilities, respectively,
{which are}
determined by
the responses of the appropriate subdetectors.}

In order to estimate the signal efficiency and 
to optimize the event selection, 
we use {Monte Carlo} (MC) samples.
The signal and the background events from generic $\tau^+\tau^-$ decays are 
 generated by KKMC/TAUOLA~\cite{KKMC}. 
In the signal MC, we generate $\tau^+\tau^-$, where 
a $\tau$ decays into  
{a lepton and an $f_0(980)$ meson} 
{according to {two-body {phase space},}}
and the other $\tau$ decays generically.
In the signal MC the $f_0(980)$  meson decays into $\pi^+\pi^-$ 
with mass and width {parameters of the $f_0(980)$}
based on the result of our study of 
$\gamma\gamma\to\pi^+\pi^-$~\cite{f0model}. 
Other {backgrounds,} including
{$B\bar{B}$ and 
$e^+e^-\to q\bar{q}$ ($q=u,d,s,c$), Bhabha 
and two-photon processes} are generated by 
EvtGen~\cite{evtgen},
BHLUMI~\cite{BHLUMI}, 
and
{AAFH~\cite{AAFH}}, respectively. 
All kinematic variables are calculated in the laboratory frame
unless otherwise specified.
In particular,
variables
calculated in the $e^+e^-$ center-of-mass (CM) system
are indicated by the superscript ``CM''.

\section{Event Selection}

%
%
{We search for $\tau^+\tau^-$ events in which 
{one} $\tau$ 
(the {signal $\tau$}) decays into {a charged}  lepton
and
two charged pions from 
{an} $f_0(980)$ meson decay}
and the other $\tau$~(the {tag $\tau$}) 
decays 
into  {a final state with} 
one charged track {and} any number of additional 
photons and neutrinos.
Candidate $\tau$-pair events are required to have 
four tracks with {zero} net charge.

%
%

{We reconstruct}
four charged tracks and any number of photons within the fiducial volume
 defined by $-0.866 < \cos\theta < 0.956$,
{where $\theta$ is 
the polar angle {relative} to 
the direction opposite to 
that of 
the {incident} $e^+$ beam in 
{the} laboratory frame.}
The transverse momentum ($p_t$) of each charged track
and {the} energy of each photon ($E_{\gamma}$) 
are 
{required to satisfy} $p_t> $ 0.1 GeV/$c$ and $E_{\gamma}>0.1$ GeV,
respectively.
{For each charged track, 
the distance of the closest point with 
respect to the interaction point 
is required to be 
less than 0.5 cm in the transverse direction 
and less than 
3.0 cm in the longitudinal direction.}

%
%

{Using the plane perpendicular to the CM
thrust axis~\cite{thrust},
which is calculated from 
the observed tracks and photon candidates,
we separate the particles in an event
into two hemispheres.
These  are referred to as the signal and 
tag sides. 
The tag side contains one charged track}
while the signal side contains three charged tracks.
We require {the} charged track
{on} the signal side 
{to be} 
identified as a lepton.
The lepton 
identification criteria are 
${\cal P}(\ell) > 0.95$ 
{and}
momentum greater than 
0.8 GeV/$c$.
The electron (muon) identification
{efficiency
is} 92\% (86\%)
while
{the probability to misidentify {a} pion
as
{an} {electron} ({a} muon)}
is below 0.2\% (2\%).
In order to take into account the emission
of {bremsstrahlung photons} from {electrons,}
{the {momenta} 
of photons within 0.05 {rad} 
of the electron tracks are added 
to the momentum of the electron candidate track.}
To reduce generic $\tau^+\tau^-$ 
and $q\bar{q}$ background events,
{we veto events with a photon in the signal side.}

%
%

To ensure that the missing particles are neutrinos rather
than photons or charged particles that pass  outside the detector acceptance,
we impose requirements on the missing 
momentum $\vec{p}_{\rm miss}$,
which 
is
calculated by subtracting the
vector sum of the momenta of all tracks and photons
from the sum of the $e^+$ and $e^-$ beam momenta.
We require that the magnitude of $\vec{p}_{\rm miss}$
be  greater than 0.8 GeV/$c$,
and {that} its direction point into the fiducial volume of the
detector.
Furthermore,
we reject the event if the  
direction of the missing momentum 
{traverses the
gap between the barrel and the ECL endcap} 
since an undetected photon 
{may fake a neutrino candidate(s).}

%
%

\begin{figure}
\begin{center}
       \resizebox{0.4\textwidth}{0.4\textwidth}{\includegraphics
        {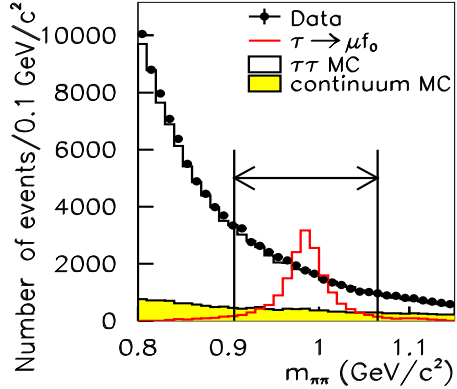}}
 \vspace*{-0.5cm}
 \caption{
 I
The invariant mass distribution of $\pi^+\pi^-$ 
in the signal region.
 While the signal MC ($\tau^-\to\mu^-f_0(980)$)
 distribution is normalized arbitrarily, 
 {the data and background MC} are normalized to the same luminosity.
 {Selected regions are indicated
 by {the}
 arrows from the marked cut {boundaries.}}}
\label{fig:f0mass}
\end{center}
\end{figure}

{The $f_0(980)$ meson is
reconstructed
from two {oppositely-charged tracks assuming a pion {mass} for both
tracks}}
in the signal side.
The charged pion from the $f_0(980)$ decay is identified by
a condition ${\cal{P}}(K/\pi) < 0.6$ for two tracks except
a lepton candidate in the signal side.
Figure~\ref{fig:f0mass} shows 
the invariant mass of $\pi^+\pi^-$ in the signal side.
{We require that
the invariant mass  satisfy {the  condition}}
{{906 MeV/$c^2 < M_{\pi^+\pi^-} < $ 1065 {MeV/$c^2$}.}}
{In order to avoid fake $f_0(980)$ candidates
from {photon} conversions
(i.e. $\gamma \rightarrow e^+e^-$),}
we require that two pion candidates 
have ${\cal{P}}(e) <0.1$.
Furthermore, we require that 
${\cal{P}(\mu)} <0.1$
to suppress two-photon $e^+e^-\mu^+\mu^-$ background events.
We require that
the cosine of the opening angle between the lepton
and {the} 
reconstructed $f_0(980)$ candidate in the CM system 
($\cos \theta^{\rm CM}_{{\rm lepton}-f_0}$)
be $0.7 <\cos \theta^{\rm CM}_{{\rm lepton}-f_0} < 0.98$.

%
%

To reject $q\bar{q}$ background,
we require that the magnitude of thrust ($T$)
be {in the range} 
0.90 $< T  <$ 0.97 (see Fig.~\ref{fig:cut_fig} (a)). 
We also require $5.29$ GeV 
$< E^{\mbox{\rm{\tiny{CM}}}}_{\rm{vis}} <$ {10.00 GeV,} 
where $E^{\mbox{\rm{\tiny{CM}}}}_{\rm{vis}}$ 
is the total visible energy in the CM system, defined as 
the sum of the energies of {the} lepton, $f_0(980)$ candidate,
the charged track in the tag side (with a pion mass hypothesis)
and all photon candidates (see Fig.~\ref{fig:cut_fig} (b)) .

To suppress the $B\bar{B}$ and $q\bar{q}$ background,
{it is required that 
the number of photons in the tag side $n_{\gamma}^{\rm{TAG}}$ 
be $n_{\gamma}^{\rm{TAG}}\leq 3$ 
if the charged track is the tag side is an electron or a muon 
{(a leptonic tag)}
and $n_{\gamma}^{\rm{TAG}}\leq 2$
if the track in the tag side
is a hadron {(a hadronic tag).}}

%

Since neutrinos are 
emitted only in the tag side,
the direction of
$\vec{p}_{\rm miss}$
should 
{be on} 
the tag side of the event.
The cosine of the
opening angle between
$\vec{p}_{\rm miss}$
and the charged track in the tag side 
{in the CM system,}
$\cos \theta^{\mbox{\rm \tiny CM}}_{\rm tag-miss}$, 
{{should be} in the range 
$0.4<\cos \theta^{\mbox{\rm \tiny CM}}_{\rm tag-miss}<0.98$
(see Fig.~\ref{fig:cut_fig} (c)). 
The reconstructed 
{mass
of a charged track (with a pion mass hypothesis) and photons}
in the tag side 
$m_{\rm tag}$, 
is required to be less than 1.00 GeV/$c^2$ (see Fig.~\ref{fig:cut_fig} (d)).
For all kinematic distributions shown in Fig.~\ref{fig:cut_fig},
reasonable agreement between the data and background MC is observed.

\begin{figure}
\begin{center}
       \resizebox{0.8\textwidth}{0.8\textwidth}{\includegraphics
 {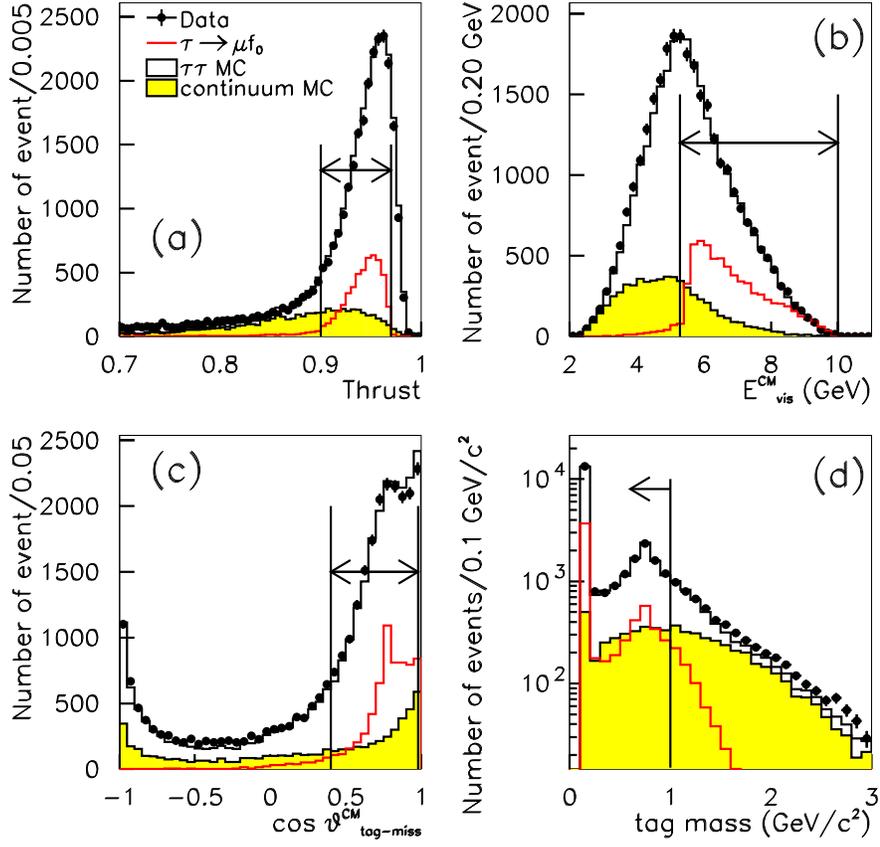}}
 \vspace*{-0.5cm}
 \caption{
 {Kinematic distributions used in the event selection:
 (a) the magnitude of the thrust vector using all track and photon
 candidates;
 (b) total energy  using all track and photon
 candidates in the CM system;
 (c) the cosine of the opening angle between a charged track in the 
 tag side and
 missing particles in the CM system ($\cos \theta_{\rm tag-miss}^{\rm CM}$);
 (d) the reconstructed mass in the 
 tag side using a charged track and photons.
 While the signal MC ($\tau^-\to\mu^-f_0(980)$)
 distribution is normalized arbitrarily, 
 {the data and background MC} are normalized to the same luminosity.
 {Selected regions are indicated
 by {the}
 arrows from the marked cut {boundaries.}}}
}
\label{fig:cut_fig}
\end{center}
\end{figure}

%
%
Finally, 
to suppress  backgrounds from generic 
$\tau^+\tau^-$ and $q\bar{q}$ events, 
we apply a selection based on the magnitude of the missing momentum ${p}_{\rm{miss}}$ 
and {the}
missing mass squared $m^2_{\rm{miss}}$.
{{We apply different selection criteria 
since the number of
emitted neutrinos is two for {the} leptonic tag
while it is one for {the} hadronic tag.}}
We require
the following relation between
$p_{\rm{miss}}$ and $m^2_{\rm{miss}}$:
$p_{\rm{miss}} > -5.0\times m^2_{\rm{miss}}$
and 
$p_{\rm{miss}} > 7.0\times m^2_{\rm{miss}}- 1.0$
for {the} hadronic tag
and 
$p_{\rm{miss}} > -7.0\times m^2_{\rm{miss}}-1.0$
and 
$p_{\rm{miss}} > 1.8\times m^2_{\rm{miss}}- 0.8$
for {the} leptonic tag,
where $p_{\rm{miss}}$ is in GeV/$c$ and
$m_{\rm{miss}}$ is in GeV/$c^2$
(see Fig. \ref{fig:pmiss_vs_mmiss2}).
While this cut retains 56\% of {the signal,}
97\% of the generic $\tau^+\tau^-$ and 96\% of $q\bar{q}$ 
{backgrounds}
are removed.

\begin{figure}
\begin{center}
 \resizebox{0.8\textwidth}{0.8\textwidth}{\includegraphics
 {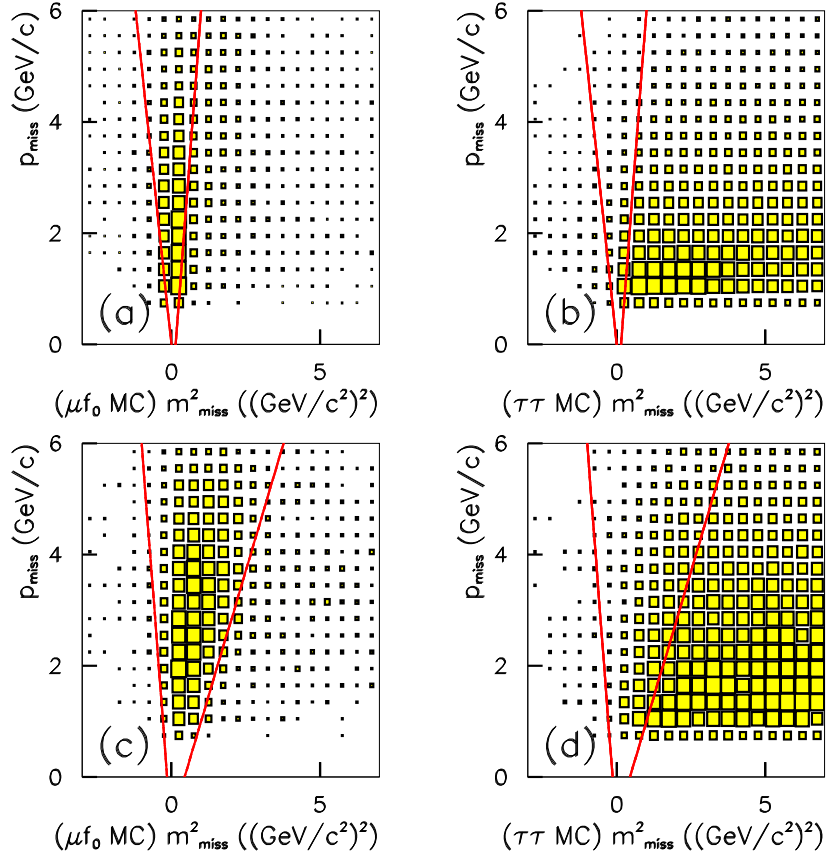}}
 \vspace*{-0.5cm}
 \caption{
Scatter-plots of
{$p_{\rm miss}$ 
{vs.} 
$m_{\rm miss}^2$:
(a) and (b)
show
the signal MC 
($\tau^-\to\mu^-f_0(980)$)
and 
the generic $\tau^+\tau^-$ MC 
distributions,
respectively,
for the hadronic tags
while (c) and
(d) show
the same distributions
for the {leptonic tags.}}
Selected regions are indicated by lines.}
\label{fig:pmiss_vs_mmiss2}
\end{center}
\end{figure}

\section{Signal and Background Estimation}

The signal candidates are examined in the 
two-dimensional {plot} of the $\ell^- f_0(980)$ invariant
mass~({$M_{ {\ell} f_0}$}) {versus} 
the difference of their energy from the 
beam energy in the CM system~($\Delta E$).
A signal event should have {$M_{{\ell} f_0}$}
close to the $\tau$-lepton mass ($m_{\tau}$) and
$\Delta E$ close to zero.
For all modes,
the {$M_{{\ell} f_0 }$ and $\Delta E$} 
resolutions are parameterized
from fits to the signal MC {distributions,}  
with  an asymmetric Gaussian function that takes into account 
{initial-state} radiation.
These {Gaussians} have widths
$\sigma^{\rm{high/ low}}_{M_{ef_0}} = 4.8/ 5.6 $ MeV/$c$$^2$ and
$\sigma^{\rm{high/ low}}_{\Delta E} = 12/ 19$ MeV
for the $\tau^-\rightarrow e^-f_0(980)$ mode{,}
and
$\sigma^{\rm{high/ low}}_{M_{{\mu f_0}}} = 5.1/ 5.3$ MeV/$c$$^2$ and
$\sigma^{\rm{high/ low}}_{\Delta E} = 13/ 19$ MeV
for the $\tau^-\rightarrow \mu^-f_0(980)$ mode,}
where the ``high/low'' superscript indicates the higher/lower side
of the peak.

\begin{figure}
\begin{center}
       \resizebox{0.5\textwidth}{0.5\textwidth}{\includegraphics
 {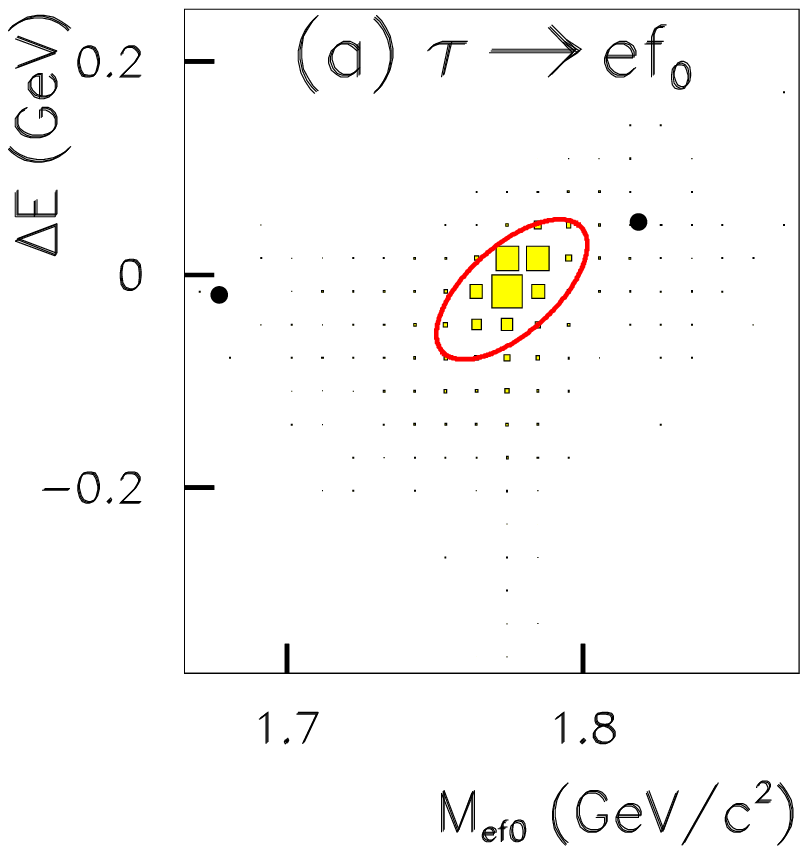}}
 \hspace*{-0.5cm}
       \resizebox{0.5\textwidth}{0.5\textwidth}{\includegraphics
 {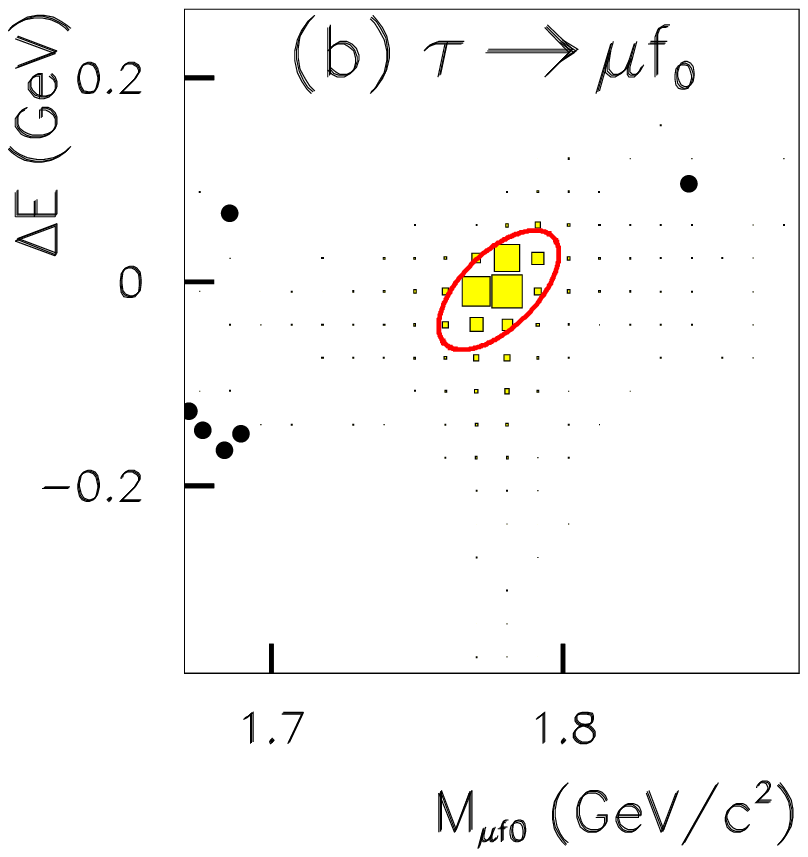}}
 \caption{
Scatter-plots in the $M_{\ell f_0}$ -- $\Delta{E}$ plane  
with
the $\pm 20 \sigma$ area for
(a) $\tau^-\rightarrow e^- f_0(980)$
and (b) $\tau^-\rightarrow \mu^- f_0(980)$.
The data are indicated by the solid circles.
The filled boxes show the MC signal distribution
with arbitrary normalization.
The elliptical signal
{regions}
shown by a solid curve
are used for evaluating the signal yield.
{No data events are observed in
the elliptical signal regions.}}
\label{fig:open}
\end{center}
\end{figure}

{{To evaluate the branching {fractions,}
we use  {elliptical signal regions}
{that} {contain} 90\%
of the MC signal {events satisfying} all {selection criteria.}
We blind the data in the signal region
{until all selection criteria are finalized}
so as not to bias our choice of selection criteria. }}
Figure~\ref{fig:open} 
shows scatter-plots
for the data and the signal MC distributed over $\pm 20\sigma$
in the {$M_{{\ell f_0}}-\Delta E$} plane.
For the $e^-f_0(980)$ mode, {{the}
dominant background comes} 
from
two-photon processes  in the $\Delta E>$ 0 GeV region,
while $q\bar{q}$ and generic $\tau^+\tau^-$ events 
are negligible 
due to 
{the} low electron fake rate.
{For the $\mu^-f_0(980)$ mode,
{the dominant sources of background}  
are 
{random} 
combinations of three pions 
from $q\bar{q}$ 
{or} generic $\tau^+\tau^-$ {processes}
{in which one pion is misidentified as {a} muon.}
The $q\bar{q}$ events are distributed 
in the $\Delta E>$ 0 GeV region
while $\tau^+\tau^-$ events 
are distributed 
in the $\Delta E<$ 0 GeV and 
$M_{\mu f_0} < $ $m_{\tau}$ region.}
We estimate the expected number of background events in the signal region
using the MC distribution with looser selection criteria and extrapolation
from the data in the sideband region defined as
the $\pm 20 \sigma$ region without the signal ellipse.
The signal efficiency and 
the number of expected background 
events for each mode 
are summarized in Table~\ref{tbl:eff}.
{After estimating the background,  we 
{open {the blinded} regions and 
{find no candidate events} {in either mode.}}}

The dominant systematic uncertainties
for this analysis
{come
from $f_0(980)$ selection
and
tracking efficiencies.}
The systematic uncertainty for $f_0(980)$ selection
is studied by changing mass and width of the $f_0(980)$
{using uncertainties from the result of our study~\cite{f0model},}
and is 9.9\% and 9.2\% 
for $e^-f_0(980)$ and $\mu^-f_0(980)$, respectively.
The uncertainty due to the charged track 
{reconstruction} is 
estimated to be 1.0\% per charged track,
therefore, the total uncertainty due to the charged track finding is 4.0\%
for both modes.
{The uncertainties due to lepton {identification} 
{are} 
2.2\% and 1.9\% for 
{electrons and muons,} respectively.}
{The uncertainty due to pion {identification} 
{is} 1.0\% per pion. }
The uncertainties due to MC statistics and luminosity
are estimated to be 
{(2.7-2.9)\%} and 1.4\%, respectively.
The uncertainty due to the trigger efficiency is negligible 
compared with the other uncertainties.
All these uncertainties are added in quadrature, 
and the total systematic uncertainties for the $e^- f_0(980)$ and $\mu^- f_0(980)$ mode are
11.5\% and
10.8\%, 
respectively.

\begin{table}
\begin{center}
\caption{ The signal efficiency($\varepsilon$), 
the number of expected background {events}  ($N_{\rm BG}$)
estimated from the  sideband data, 
{total} 
systematic uncertainty  ($\sigma_{\rm syst}$),
{the} number of observed events 
in the signal region ($N_{\rm obs}$), 
90\% C.L. upper limit on the number of signal events including 
systematic uncertainties~($s_{90}$) 
and 90\% C.L. upper limit on 
the branching  fraction including 
the branching fraction $f_0(980)\to\pi^+\pi^-$~($\cal{UL}$)
for each individual mode. }
\label{tbl:eff}
\begin{tabular}{c|cccccc}\hline \hline
Mode &  $\varepsilon$~{(\%)} & 
$N_{\rm BG}$  & $\sigma_{\rm syst}$~{(\%)}
& $N_{\rm obs}$ & $s_{90}$ & 
${\cal{UL}}(10^{-8})$ \\ \hline
$\tau^-\to e^- f_0(980)$ & 5.80 & $0.10\pm{0.07}$ & 
11.5 & 
0 & 2.41  & 3.4\\
$\tau^-\to\mu^- f_0(980)$ & 6.02 & $0.11\pm{0.08}$ & 
10.8 & 0  & 2.40  & 3.2 \\
\hline\hline
\end{tabular}
\end{center}
\end{table}

\section{Upper Limits on the branching fractions}

We set upper limits on the branching fractions 
of $\tau^-\to\ell^- f_0(980)$
based on the Feldman-Cousins method~\cite{cite:FC}.
The 90\% C.L. upper limit on the number of signal events 
including  a systematic uncertainty~($s_{90}$) is obtained 
{using} the POLE program without conditioning~\cite{pole}
{based on}
the number of expected {background events}, 
{the number of observed events}
and the systematic uncertainty.
The upper limit on the branching fraction ($\cal{B}$) is then given by
\begin{equation}
{{\cal{B}}(\tau^-\to\ell^-f_0(980))
\times
{{\cal{B}}(f_0(980)\to\pi^+\pi^-) <
\displaystyle{\frac{s_{90}}{2N_{\tau\tau}\varepsilon{}}}}},
\end{equation}
where $N_{\tau\tau}$ is the number of $\tau^+\tau^-$pairs, and 
$\varepsilon$ is the signal efficiency.
{The value {$N_{\tau\tau} =  616.6\times 10^6$}} is obtained 
from 
{the} integrated luminosity times 
the cross section of {$\tau$-pair production,} which 
is calculated 
{in the updated version of 
KKMC~\cite{tautaucs} to be 
$\sigma_{\tau\tau} = 0.919 \pm 0.003$ nb.}
We set the following 90\% C.L. upper limits on the 
branching fraction {products:} 
${\cal{B}}(\tau^-\rightarrow e^-f_0(980))\times
{\cal{B}}(f_0(980)\rightarrow \pi^+\pi^-)
 < 3.2\times 10^{-8}$
and 
${\cal{B}}(\tau^-\rightarrow \mu^-f_0(980))\times 
{\cal{B}}(f_0(980)\rightarrow \pi^+\pi^-) < 3.4\times 10^{-8}$.

\section{Summary}
We have searched for {lepton-flavor-violating} $\tau$ decays 
into {a} lepton and $f_0(980)$ {meson} 
using 671 fb$^{-1}$ of data.
No events are observed and
we set  the following 90\% C.L. upper limits 
on the branching fraction 
{products:} 
${\cal{B}}(\tau^-\rightarrow e^-f_0(980))\times
{\cal{B}}(f_0(980)\rightarrow \pi^+\pi^-)
 < 3.2\times 10^{-8}$
and 
${\cal{B}}(\tau^-\rightarrow \mu^-f_0(980))\times 
{\cal{B}}(f_0(980)\rightarrow \pi^+\pi^-) < 3.4\times 10^{-8}$.
This is the first search for these {$\tau$ LFV decay} modes.

\section*{Acknowledgments}

We are grateful to M.J.~Herrero for stimulating discussions.
We thank the KEKB group for the excellent operation of the
accelerator, the KEK cryogenics group for the efficient
operation of the solenoid, and the KEK computer group and
the National Institute of Informatics for valuable computing
and SINET3 network support as well as Tau lepton physics research 
center of Nagoya University. We acknowledge support from
the Ministry of Education, Culture, Sports, Science, and
Technology of Japan and the Japan Society for the Promotion
of Science; the Australian Research Council and the
Australian Department of Education, Science and Training;
the National Natural Science Foundation of China under
contract No.~10575109 and 10775142; the Department of
Science and Technology of India; 
the BK21 program of the Ministry of Education of Korea, 
the CHEP src program and Basic Research program (grant 
No. R01-2008-000-10477-0) of the 
Korea Science and Engineering Foundation;
the Polish State Committee for Scientific Research; 
the Ministry of Education and Science of the Russian
Federation and the Russian Federal Agency for Atomic Energy;
the Slovenian Research Agency;  the Swiss
National Science Foundation; the National Science Council
and the Ministry of Education of Taiwan; and the U.S.\
Department of Energy. This work is supported by
a Grant-in-Aid for Science Research on Priority Area (New 
Development of Flavor Physics) from the Ministry of Education, 
Culture, Sports, Science and Technology of Japan and Creative 
Scientific Research (Evolution of Tau-lepton Physics) from the 
Japan Society for the Promotion of Science.


\begin{thebibliography}{99}

\bibitem{brig} A.~Brignole and A.~Rossi, Nucl. Phys. B {\bf 701},3 (2004). 
\bibitem{arg} E.~Arganda, M.J.~Herrero and J.~Portoles,
JHEP 0806:079, 2008.

\bibitem{lf0}
        C.-H.~Chen and C.-Q.~Geng,
        Phys.~Rev.~D {\bf 74}, 035010 (2006).

\bibitem{higgsLFV}
K.~S.~Babu and C.~Kolda,
Phys.~Rev.~Lett. {\bf 89}, 241802 (2002);
A.~Dedes, J.~R.~Ellis and M.~Raidal,
Phys.~Lett.~B {\bf 549}, 159 (2002);
M.~Sher,
Phys.~Rev.~D {\bf 66}, 057301 (2002).



\bibitem{leta}
Y.~Miyazaki {\it et al.} (Belle Collaboration),
Phys.\ Lett.\ B {\bf 648}, 341 (2007);
B.~Aubert {\it et al.} (BaBar Collaboration),
Phys.\ Rev.\ Lett. {\bf 98}, 061803 (2007).

\bibitem{lll}
Y.~Miyazaki {\it et al.} (Belle Collaboration),
Phys.\ Lett.\ B {\bf 660}, 154 (2007);
B.~Aubert {\it et al.} (BaBar Collaboration),
Phys.\ Rev.\ Lett. {\bf 99}, 251803 (2007).



\bibitem{kekb}
S.~Kurokawa and  E.~Kikutani, Nucl. Instr. {and} Meth. A {\bf 499}, 1
(2003), and other papers included in this Volume.



\bibitem{Belle}
A.~Abashian {\it et al.} (Belle Collaboration),
Nucl. Instr. and Meth. A {\bf 479}, 117 (2002).


\bibitem{EID}
        K.~Hanagaki {\it et al.},
        Nucl. Instr. and Meth.  A {\bf 485}, 490 (2002).



\bibitem{MUID}
	{A.~Abashian {\it et al.}},
        Nucl. Instr. and Meth. A 
	{\bf 491}, 69 (2002).




\bibitem{KKMC}
        S.~Jadach {\it et al.},
        Comp. Phys. Commun. {\bf 130}, 260 (2000).

\bibitem{f0model}
        T.~Mori {\it et al.} (Belle Collaboration),
        Phys.~Rev.~D {\bf 75}, 051101 (2007).

\bibitem{evtgen}
	D.~J.~Lange,
        {Nucl. Instr. and Meth. A} 
	{\bf 462}, 152 (2001).




\bibitem{BHLUMI}
        S.~Jadach {\it et al.},
        Comp. Phys. Commun. {\bf 70}, 305 (1992).


\bibitem{AAFH}
        F.~A.~Berends {\it et al.},
        Comp. Phys. Commun. {\bf 40}, 285 (1986).


\bibitem{thrust}
S.~Brandt {\it et al.},
        Phys.\ Lett.\ {\bf 12}, 57 (1964);
E.~Farhi,
        Phys.\ Rev.\ Lett.\ {\bf 39}, 1587 (1977).




\bibitem{cite:FC}
        G.~J.~Feldman and {R.~D. Cousins,}
        Phys.\ Rev.\ D {\bf 57}, 3873 (1998).



\bibitem{pole}
       See {http://www3.tsl.uu.se/${}^{\sim}$conrad/pole.html;}
        J.~Conrad {\it et al.},
        Phys.\ Rev.\ D {\bf 67}, 012002 (2003).    


\bibitem{tautaucs}
        S.~Banerjee {\it et al.},
        Phys.\ Rev.\ D {\bf 77}, 054012 (2008).



\end{thebibliography}
\end{document}